\documentclass[12pt]{article}
\usepackage{graphicx}
\usepackage{natbib}
\usepackage{journals}

\def\fracd#1#2{{\displaystyle\frac{#1}{#2}}}

\begin{document}

\title{The current best estimate of the Galactocentric distance of the Sun based on comparison of different statistical techniques}
\author{Zinovy Malkin$^{1,2}$}
\date{$^1$Pulkovo Observatory, St.~Petersburg 196140, Russia\\
  $^2$St. Petersburg State University, St.~Petersburg 198504, Russia\\
  e-mail: malkin@gao.spb.ru}
\maketitle

\begin{abstract}
In this paper, the current best estimate of a fundamental Galactic parameter Galactocentric distance of the Sun $R_0$ has been evaluated
using all the available estimates published during the last 20 years.
Unlike some other studies, our analysis of these results showed no statistically significant trend in $R_0$ during this period.
However we revealed a statistically valuable improvement in the $R_0$ uncertainties with time of about 0.2 kpc for 20~years.
Several statistical techniques have been used and compared to obtain the most reliable common mean of 52 determinations made in 1992--2011 and
its realistic uncertainty.
The statistical methods used include unweighted mean, seven variants of the weighted mean, and two variants of median technique.
The $R_0$ estimates obtained with these methods range from 7.91 to 8.06 kpc with uncertainties varying from 0.05 to 0.08 kpc.
The final value derived in this analysis is $R_0 = 7.98 \pm 0.15\,|_{stat} \pm 0.20\,|_{syst}$ kpc, which can be recommended as the current
best estimate of the Galactocentric distance of the Sun.
For most of the practical applications the value of $R_0 = 8.0 \pm 0.25$ kpc can be used.
\end{abstract}

\section{Introduction}

An accurate knowledge of the distance from the Sun to the center of the Galaxy, $R_0$, is important in many fields of astronomy and space sciences
such as establishing a distance scale in the Universe, studying the structure and dynamics of the Galaxy, determination of the distance to
extragalactic objects, etc.
In particular, the primary motivation for this study was a wish to improve the accuracy of modeling of the Galactic aberration \citep{Malkin2011fe}.

During decades, many tens of scientifically meaningful determinations of $R_0$ were made.
Various methods and observational data were used to obtain these results, which sometimes substantially differ from each other.
The uncertainties associated with this determinations also vary in wide range, and not always correspond to the real accuracy of reported results,
which was not always properly assessed.
For these reasons, it may be difficult for the user to choose the best $R_0$ value suitable for his application.
These shortcomings of the original measurements can be mitigated in an average value or a common mean (CM), which is a generally used procedure.
Equally important is to derive a realistic uncertainty associated with this combined value.
Both underestimated and overestimated uncertainties are undesirable.

Table~\ref{tab:means} shows the latest average estimates of $R_0$.
All the three values are statistically consistent, i.e. lies inside the overlapping 1$\sigma$ intervals.
However, the uncertainty of the first result is too large to satisfy the need of many exacting applications.
The reason may be that \citet{Kerr1986} computed the uncertainty of the final result as the standard deviation (STD) of the individual
determinations with respect to the unweighted mean value, and ignored statistical error equal to 0.22.
On the other hand, the uncertainty of the results of \citet{Nikiforov2004} and \citet{Avedisova2005} may be underestimated because it seems
to be just a formal error coming from the least squares procedure without a check of consistency of the data used in the computation.
Notice that the result of \citet{Avedisova2005} is less precise because, most probably, it is obtained from relatively small set of data with
respect to the work of \citet{Nikiforov2004}.
The result of \citet{Reid1993} based on a comprehensive analysis of many data looks as the most realistic both with respect to $R_0$ value
and its uncertainty.
However it was obtained about 20~years ago, and evidently needs revision.
In should be noted that all these results were obtained with detailed considerations of the physical aspects of observational aspects, but maybe without
sufficient statistical analysis of the published estimates.
In particular, they used simple (un)weighted mean to compute a statistical estimate of CM without accounting for possible statistical discrepancy
of the published data.

\begin{table}
\centering
\small
\caption{Previous average estimates of $R_0$}
\label{tab:means}
\begin{tabular}{lcc}
\hline
\multicolumn{1}{c}{Paper} & Period covered & $R_0$, kpc \\
\hline
\citet{Kerr1986}      & 1974--1986 & $8.5 \pm 1.1~~$ \\
\citet{Reid1989}      & 1974--1987 & $7.7 \pm 0.7~~$ \\
\citet{Reid1993}      & 1974--1992 & $8.0 \pm 0.5~~$ \\
\citet{Nikiforov2004} & 1974--2003 & $7.9 \pm 0.2~~$ \\
\citet{Avedisova2005} & 1992--2005 & $7.8 \pm 0.32$ \\
\hline
\end{tabular}
\end{table}

A goal of this study is to derive a new best estimate of $R_0$.
We will use all available determinations published during last 20~years (it will be shown below that earlier results
do not contribute substantially to the final result).
Besides we will use several statistical techniques which allow processing not only consistent but also discrepant data.
This task is not unique.
It is well known metrological problem actual for many fields of science and engineering \citep{Cox2004}.
In particular, in this study we consider several methods developed for evaluation of the best estimates of some physical constants
\citep{MacMahon2004,Chen2011}.

This paper is limited in scope to the statistical analysis of the $R_0$ determinations found in the literature.
We will not attempt to discuss the methods involved in the various determinations of $R_0$, either theoretical or observational.
This study is primarily aimed at testing of a capability of different mathematical approaches in deriving the best current estimates of the Galactic
fundamental constants.

The paper is organized as follows.
Section~\ref{sect_methods} described statistical methods used to evaluate the best estimate of $R_0$.
Section~\ref{sect_input} presents all the available $R_0$ estimates published during last 20~years and their preliminary analysis.
The methods described in Section~\ref{sect_methods} will be used in Section~\ref{sect_compute} to obtain the final result.
The work is closed with Concluding remarks.

\section{Statistical methods used}
\label{sect_methods}

Let we have $n$ measured values of a physical quantity $x_i$ with associated reported STD (uncertainties) $s_i$, $i=1 \ldots n$.
Our goal is to compute the best estimate of this quantity and its associated uncertainty.
This task is known as the problem of computation of a common mean (CM).
A solution of the problem is usually not so simple in practice.
Input values $x_i$ generally come from various research groups and are obtained with different methods, observational techniques and data sets.
Also the uncertainties $s_i$ are, as a rule, obtained not in uniform way, and their statistical properties are usually not known.
These circumstances may make input data rather discrepant than consistent, and using simple statistical methods from textbooks on the subject like
most often used classical weighted mean may be not justified.
For these reason, several alternative statistical methods were developed, see, e.g., \citet{MacMahon2004,Chen2011,Malkin2011g} and papers cited therein.
For evaluation of $R_0$, several of these methods were selected that are most representative and relevant, in our opinion, to this study.
They are briefly discussed in this section.

{\bfseries Weighted mean (WM)}.
This classical method is most widely used.
The mean value is computed by
\begin{equation}
\bar{x}_w = \fracd{\sum_{i=1}^n  p_i x_i}{p} \,, \\
\label{eq:wm_mean}
\end{equation}
with STD
\begin{equation}
\sigma_{w1} = \fracd{1}{\sqrt{p}} \,,
\label{eq:wm_sigma_1}
\end{equation}
using weights
\begin{equation}
p_i = \fracd{1}{s_i^2} \,, \quad p = \sum_{i=1}^n {p_i} \,.
\label{eq:weights}
\end{equation}

The same formulae (\ref{eq:wm_mean}) and (\ref{eq:wm_sigma_1}) are used in computation of CM is also used in the LRW, NR, and MP methods described below.
However, these methods use modified procedures to compute the weights of the input measurements.

Another important statistics is the reduced (normalized) $\chi^2$ computed by
\begin{equation}
\frac{\chi^2}{dof} = \frac{\sum_{i=1}^n {p_i (x_i - \bar{x}_w)^2}}{n-1} \,.
\label{eq:chi2}
\end{equation}

It is close to unity for consistent measurements, and is much greater than 1 in case of discrepant data or underestimated reported uncertainties.

Least squares approach leads to alternative estimate of the WM uncertainty.
The solution of the least square problem $x_i=\bar{x}_w+\varepsilon_i$ with weights $p_i$ gives the same WM value $\bar{x}_w$, but
another estimate of its uncertainty:
\begin{equation}
\sigma_{w2} = \sqrt{\fracd{\sum_{i=1}^n {p_i (x_i - \bar{x}_w)^2}}{p\,(n-1)}} \,.
\label{eq:wm_sigma_2}
\end{equation}

One can see that $\sigma_{w1}$, depends on $s_i$ and does not depend on the scatter of $x_i$, whereas $\sigma_{w2}$ depends on both
{\textsl relative} values of $s_i^2$ and the scatter of $x_i$.
They are equal to each other in the case of consistent data, i.e. when $\chi^2/dof$ = 1.
Difference between the two $\sigma$ estimates may be attributed to systematic errors in $x_i$ and/or underestimating of $s_i$.
It's important to note that neither $\sigma_{w1}$ nor $\sigma_{w2}$ use all the input information, namely scatter of the input measurements
and absolute and relative values of their uncertainties.
After investigation of behavior of these estimates using simulated and real data, \citet{Malkin2001k} proposed the combined estimate:
\begin{equation}
\sigma_{w3} = \sqrt{\sigma_{w1}^2+\sigma_{w2}^2} \ .
\label{eq:sigma_3}
\end{equation}

The latter approach can provide more stable and realistic uncertainty estimate in practical cases, see, e.g., \citep{Malkin2001m}.
More detailed discussion on these three approaches to compute the WM STD and several simulated and practical examples are given in \citet{Malkin2011g}.

Another procedure to compute the WM uncertainty was proposed by \citet{Gray1990}.
It uses Bayesian approach and is recommended by the authors for evaluation of experimental data in case of doubt about correctness of reported uncertainties.
In this method, the CM is also computed by (\ref{eq:wm_mean}), but its STD is given by
\begin{equation}
\sigma_{w4} = \sqrt{\fracd{\sum_{i=1}^n {p_i (x_i - \bar{x_w})^2}}{p\,(n-3)}} \,.
\label{eq:sigma_4}
\end{equation}

One can see that this approach is similar to the least square one, but gives a bit more conservative estimate of the uncertainty.

{\bfseries Unweighted mean (UWM).}
The UWM estimate can be justified when the uncertainties of the input measurements are not given, not trusted or suspected to be not consistent.
This may be also the case for $R_0$ determinations, see e.g. \citet{Kerr1986}.
Unweighted CM estimate and its uncertainty is computed by
\begin{equation}
\bar{x}_u = \fracd{\sum_{i=1}^n x_i}{n} \,,
\label{eq:uwm_m}
\end{equation}
\begin{equation}
\sigma = \sqrt{\fracd{\sum_{i=1}^n (x_i - \bar{x}_u)^2}{n(n-1)}} \,,
\label{eq:uwm_s}
\end{equation}
which corresponds to $\sigma_{w2}$ computed with unit weights.

{\bfseries Limitation of relative weights (LRW).}
This procedure is adopted by the Nuclear Data Section of International Atomic Energy Agency \citep{Nichols2004}.
It is developed to prevent a measurements with very small reported uncertainty from dominating CM.
According to this method, no input value should have a relative weight $p_i/p$ greater than 0.5.
When such a value found, its uncertainty should be increased, and then a weighted mean should be recomputed.

{\bfseries Normalized residuals (NR).}
This method introduced by \citet{James1992} is also intended to limit a relative statistical weights of input values.
However, unlike LRW method, the uncertainties of only discrepant data are adjusted.
For this purpose, the normalized residuals $r_i$ are computed by
\begin{equation}
r_i = \sqrt{\fracd{p\,p_i}{p-p_i}} (x_i - \bar{x}_w) \,.
\label{eq:norm_res}
\end{equation}

The limiting value of $r_i$ is defined as
\begin{equation}
r_0 = \sqrt{1.8\ln n + 2.6} \,.
\label{eq:norm_res_lim}
\end{equation}

For any measurement with $|r_i|>r_0$, the uncertainty of the i-th measurement is inflated.
The procedure is started from the measurement with the largest $|r_i|$, and repeated until no $r_i$ will be greater than $r_0$.

{\bfseries Mandel--Paule (MP).}
The MP estimate of CM \citep{Paule1982} is defined as $x_w$ computed by (\ref{eq:wm_mean}) with weights
\begin{equation}
w_i = \fracd{1}{s_i^2+s_b^2} \,,
\label{eq:mp_weights}
\end{equation}
where $s_b^2$ is computed by solving the equation
\begin{equation}
\sum_{i=1}^n {w_i (x_i - \bar{x}_w)^2} = n-1 \,.
\label{eq:mp_equation}
\end{equation}

This method treats the set of systematic errors of the input measurements as a source of random variability,
and $s_b^2$ is an estimate of the ``inter-laboratory'' variance.
In fact, this method simply increases original uncertainties by an additive variance to make the normalized $\chi^2/dof$ equal to unity.
So, its use is only justified in a case of the original $\chi^2/dof > 1$.
In this respect the MP approach is similar to \citet{Malkin2001k,Malkin2011g}.
\citet{Rukhin1998} defined a modified MP procedure to be as previously with $p$ instead of $(p-1)$ on the right side of (\ref{eq:mp_equation}).
The modified approach produced practically the same result in our case.

{\bfseries Median (M).}
Another approach routinely used to get the estimate of CM is computation of a median $\bar{x}_m$.
The median is known as a robust statistics less influenced by outliers.
However its standard definition does not provide an estimate of error of a median value (it makes it immune to unreliable uncertainties though).
A possible approach to compute a median uncertainty can be found in \citet{Muller1995,Muller2000a}.
Let $\bar{x}_m$ be the median of $x_i$, i.e. $\bar{x}_m = med\{x_i\}$.
Now we can compute the median of the absolute deviations (MAD) as
\begin{equation}
MAD = med\{|x_i - \bar{x}_m|\} \,.
\label{eq:mad}
\end{equation}

The uncertainty of $\bar{x}_m$ is then taken as
\begin{equation}
\sigma_m = \fracd{1.8582}{\sqrt{n-1}} \; MAD \,.
\label{eq:sigma_m}
\end{equation}

One can see that this estimate of the median uncertainty depends only on the data scatter and not on the input uncertainties.
Later \citet{Muller2000b} proposed a method to take account of the uncertainties in input data and thus compute weighted median and its STD.
However its practical realization, as pointed out by the author, is more cumbersome, and the testing results and discussion given therein
do not show clear advantage of using weighted median.

{\bfseries Bootstrap median (BM).}
The bootstrap method is based on a Monte-Carlo procedure to estimate CM and its uncertainty.
A random sample (with replacement) of length $n$ is selected from input data, and the median $x_{m,j}$ is computed.
After repeating the sampling $M$ times ($j=1 \ldots M$), the best estimate of CM and its STD are computed by
\begin{equation}
\bar{x}_b = \fracd{\sum_{j=1}^M x_{m,j}}{M} \,,
\label{eq:bm_m}
\end{equation}
\begin{equation}
\sigma_b = \sqrt{\fracd{\sum_{j=1}^M (x_{m,j} - \bar{x}_b)^2}{M-1}} \,.
\label{eq:bm_s}
\end{equation}

To obtain a reliable estimates of $\bar{x}_b$ and $\sigma_b$, sufficiently large number of sampling should be made.
After testing, we found that $M=10,000$ provides stable result not changing substantially with increasing $M$.

Note that three methods used in this work, UWM, M, and BM, do not use the uncertainties reported with data.

\section{Input $\mathbf{R_0}$ determinations}
\label{sect_input}

In this section we strived to collect all the results of determination of $R_0$ published during last 20 years.
This interval was chosen because test computations with earlier data showed no significant difference with final result of this paper
presented in Section~\ref{sect_compute}.
It will be shown below that the average result converges to a stable value during the period considered in this work.
The reader interested in the historical data can refer to, e.g., \citet{Kerr1986,Reid1993,Surdin1999,Nikiforov2004}.

All the results found in the literature were used in this work with few exceptions.
We did not use \citet{Glushkova1998} later revised by \citet{Glushkova1999}, \citet{Eisenhauer2003} revised by \citet{Eisenhauer2005},
and \citet{Paczynski1998} revised by \citet{Stanek2000}.

When both random (statistical) and systematic uncertainties are given, they were summed in quadrature.
See Table~\ref{tab:stat-syst} for details.

If two different values are given for the lower and upper boundaries of the confidence interval \citep{Ghez2008,Reid2009b,Vanhollebeke2009},
the mean value of these boundaries was used as the uncertainty of the result.
When the authors give several estimates for $R_0$ without final conclusion, an unweighted average of this estimates was computed.
Details are given in Table~\ref{tab:averaged}.

\begin{table}
\centering
\small
\caption{$R_0$ determinations with estimation of both statistical and systematic errors}
\label{tab:stat-syst}
\begin{tabular}{lc}
\hline
\multicolumn{1}{c}{Paper} & $R_0$, kpc \\
\hline
\citet{Nishiyama2006}   & $R_0 = 7.52 \pm 0.10\,|_{stat} \pm 0.35\,|_{syst}$ \\
\citet{Groenewegen2008} & $R_0 = 7.94 \pm 0.37\,|_{stat} \pm 0.26\,|_{syst}$ \\
\citet{Trippe2008}      & $R_0 = 8.07 \pm 0.32\,|_{stat} \pm 0.13\,|_{syst}$ \\
\citet{Gillessen2009a}  & $R_0 = 8.33 \pm 0.17\,|_{stat} \pm 0.31\,|_{syst}$ \\
\citet{Gillessen2009b}  & $R_0 = 8.28 \pm 0.15\,|_{stat} \pm 0.29\,|_{syst}$ \\
\citet{Matsunaga2009}   & $R_0 = 8.24 \pm 0.08\,|_{stat} \pm 0.42\,|_{syst}$ \\
\citet{Sato2010}        & $R_0 = 8.3\phantom{3} \pm 0.46\,|_{stat} \pm 1.0\phantom{3}\,|_{syst}$ \\
\hline
\end{tabular}
\end{table}

\begin{table}
\centering
\small
\caption{Averaged $R_0$ values}
\label{tab:averaged}
\begin{tabular}{lcc}
\hline
\multicolumn{1}{c}{Paper} & \multicolumn{2}{c}{Original $R_0$ estimates, kpc} \\
\hline
\citet{Glass1995}       & $8.7  \pm 0.7$  & $8.9  \pm 0.7$ \\
\citet{Layden1996}      & $7.6  \pm 0.4$  & $8.3  \pm 0.5$ \\
\citet{Genzel2000}      & $7.9  \pm 0.85$ & $8.2  \pm 0.9$ \\
\citet{Groenewegen2005} & $8.6  \pm 0.7$  & $8.8  \pm 0.4$ \\
\citet{Shen2007}        & $7.95 \pm 0.62$ & $8.25 \pm 0.79$ \\
\citet{Ghez2008}        & $7.96 \pm 0.63$ & $8.36 \pm 0.37$ \\
\citet{Majaess2009}     & $7.7  \pm 0.7$  & $7.8  \pm 0.6$ \\
\hline
\end{tabular}
\end{table}

\begin{table}
\centering
\small
\caption{$R_0$ estimates used in this study}
\label{tab:allr0}
\def\arraystretch{0.85}
\begin{tabular}{lll}
\hline
\hline
\multicolumn{1}{c}{$R_0$, kpc} & \multicolumn{1}{c}{STD} & \multicolumn{1}{c}{Reference} \\
\hline
7.9  & 0.8   & \citet{Merrifield1992} \\
8.1  & 1.1   & \citet{Gwinn1992} \\
7.6  & 0.6   & \citet{Moran1993} \\
7.6  & 0.4   & \citet{Maciel1993} \\
8.09 & 0.3   & \citet{Pont1994} \\
7.5  & 1.0   & \citet{Nikiforov1994} \\
7.0  & 0.5   & \citet{Rastorguev1994} \\
8.8  & 0.5   & \citet{Glass1995} \\
7.1  & 0.5   & \citet{Dambis1995} \\
8.3  & 1.0   & \citet{Carney1995} \\
8.21 & 0.98  & \citet{Huterer1995} \\
7.95 & 0.4   & \citet{Layden1996} \\
7.55 & 0.7   & \citet{Honma1996} \\
8.1  & 0.4   & \citet{Feast1997a} \\
8.5  & 0.5   & \citet{Feast1997c} \\
7.66 & 0.54  & \citet{Metzger1998} \\
8.1  & 0.15  & \citet{Udalski1998} \\
7.1  & 0.4   & \citet{Olling1998} \\
8.51 & 0.29  & \citet{Feast1998} \\
8.2  & 0.21  & \citet{Stanek1998} \\
8.6  & 1.0   & \citet{Surdin1999} \\
7.4  & 0.3   & \citet{Glushkova1999} \\
7.9  & 0.3   & \citet{McNamara2000a} \\
8.67 & 0.4   & \citet{Stanek2000} \\
8.2  & 0.7   & \citet{Nikiforov2000} \\
8.24 & 0.42  & \citet{Alves2000} \\
8.05 & 0.6   & \citet{Genzel2000} \\
8.3  & 0.3   & \citet{Gerasimenko2004} \\
7.7  & 0.15  & \citet{Babusiaux2005} \\
8.01 & 0.44  & \citet{Avedisova2005} \\
7.62 & 0.32  & \citet{Eisenhauer2005} \\
8.7  & 0.6   & \citet{Groenewegen2005} \\
7.2  & 0.3   & \citet{Bica2006} \\
7.52 & 0.36  & \citet{Nishiyama2006} \\
8.1  & 0.7   & \citet{Shen2007} \\
7.4  & 0.3   & \citet{Bobylev2007} \\
7.94 & 0.45  & \citet{Groenewegen2008} \\
8.07 & 0.35  & \citet{Trippe2008} \\
8.16 & 0.5   & \citet{Ghez2008} \\
8.33 & 0.35  & \citet{Gillessen2009a} \\
8.7  & 0.5   & \citet{Vanhollebeke2009} \\
7.58 & 0.40  & \citet{Dambis2009} \\
7.2  & 0.3   & \citet{Bonatto2009} \\
8.4  & 0.6   & \citet{Reid2009a} \\
7.75 & 0.5   & \citet{Majaess2009} \\
7.9  & 0.75  & \citet{Reid2009b} \\
8.24 & 0.43  & \citet{Matsunaga2009} \\
8.28 & 0.33  & \citet{Gillessen2009b} \\
7.7  & 0.4   & \citet{Dambis2010} \\
8.1  & 0.6   & \citet{Majaess2010} \\
8.3  & 1.1   & \citet{Sato2010} \\
7.80 & 0.26  & \citet{Ando2011} \\
\hline
\end{tabular}
\end{table}

\begin{figure}
\centering
\resizebox{0.7\hsize}{!}{\includegraphics[clip]{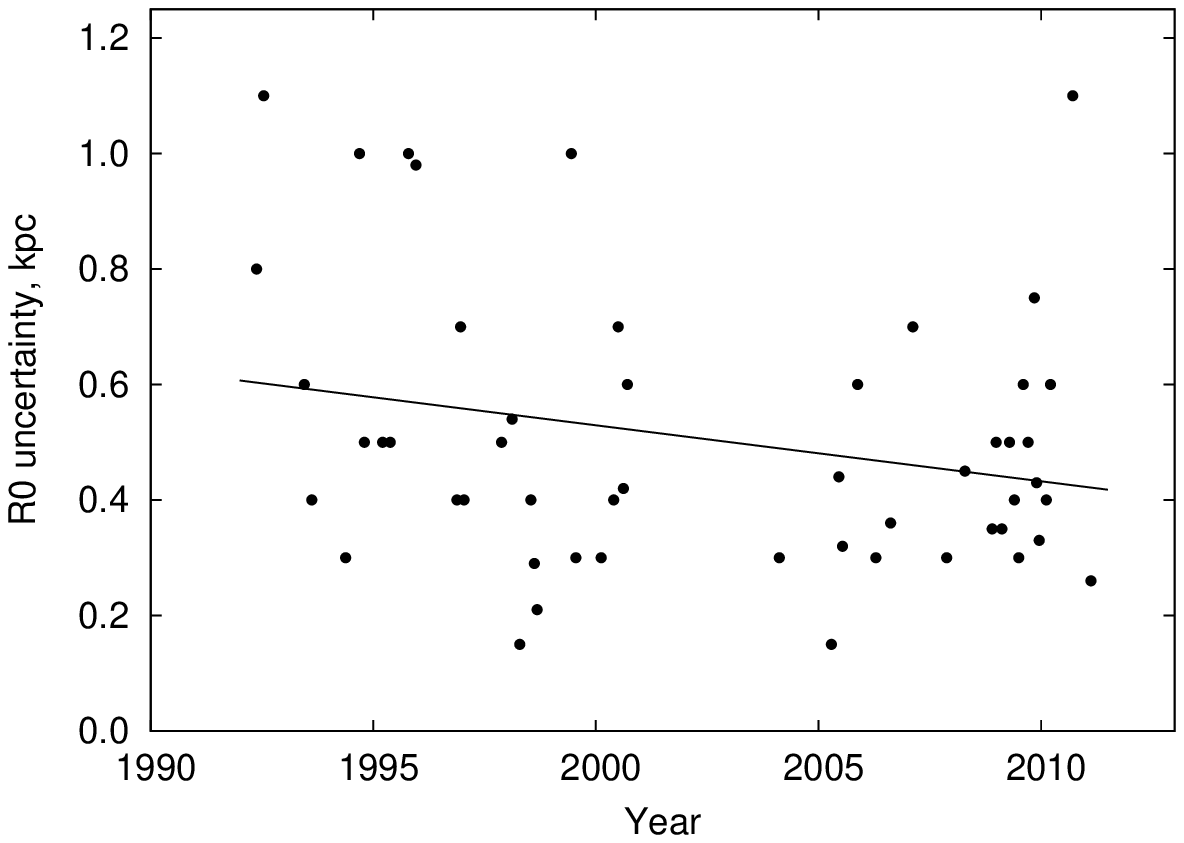}}
\caption{Uncertainty of $R_0$ estimates used in this study. Unit: kpc.}
\label{fig:data_R0_err}
\end{figure}

The first impression of the input data is a large spread in the reported uncertainties from 0.15 to 1.1~kpc with about 1.5 times improvement
during last 20~years (Fig~\ref{fig:data_R0_err}).
The value of this trend $-0.0097 \pm 0.0054$~kpc/yr is statistically significant.
A reason for such large scatter in uncertainty, except the real difference in the accuracy of the methods used, can be the fact that some authors
may be not critical enough in estimating the error of the result.
Besides, as mentioned above, only few authors give an estimate of systematic error of the reported result in addition to the usually used statistical
(formal) error.
All these factors can indicate discrepancy between determinations $R_0$, which requires careful statistical processing instead of a straightforward
computation of a weighted mean.
In particular, for this reason, \citet{Kerr1986} proposed to consider unweighted mean value for the best estimate of Galactic constants, including $R_0$.

It is important for further analysis to check the absence of the significant systematic drift in input measurements.
Many authors, e.g., \citet{Reid1993,Surdin1999,Nikiforov2004} and \citet{Foster2010} discussed the long-term trends in $R_0$ estimates.
\citet{Reid1993,Surdin1999} and \citet{Nikiforov2004} showed the large negative slope in the 1970s to the beginning of the 1990s,
As to the last 20~years, \citet{Surdin1999} showed practically no drift for 1990--1998,
\citet{Nikiforov2004} showed zero or maybe small positive drift for 1990--2003,
and \citet{Foster2010} revealed the large positive slope for 1992-2010.

\begin{figure}
\centering
\resizebox{0.7\hsize}{!}{\includegraphics[clip]{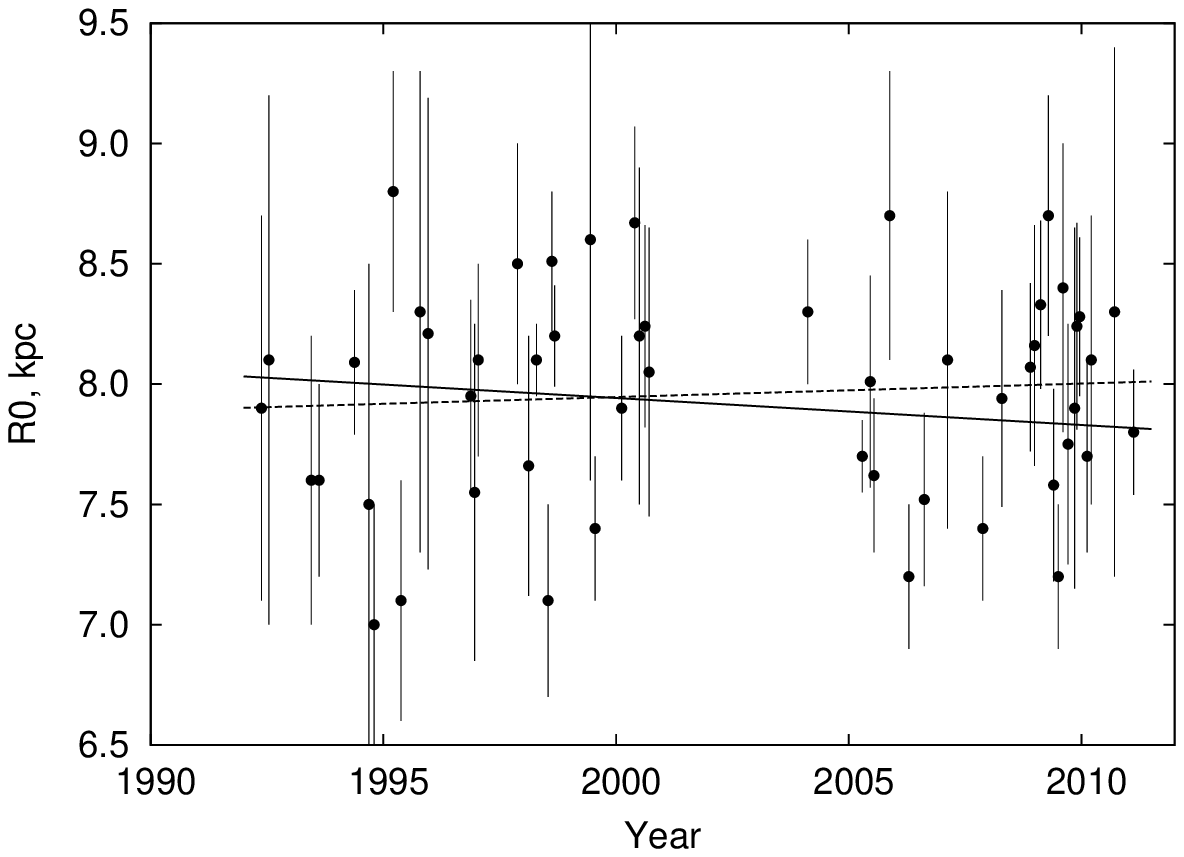}}
\caption{$R_0$ estimates used in this study. The weighted (solid line) and unweighted (dashed line) trend is also shown. Unit: kpc.}
\label{fig:data_R0}
\end{figure}

To clarify this contradiction we repeated the computation of the trend in $R_0$ estimates during the last 20 years using the full data set
considered in this paper.
All the values listed in Table~\ref{tab:allr0} are depicted in Fig.~\ref{fig:data_R0} with the epochs computed with a fraction of the year based
on the publication date taken from ADS or journal number.
When the month is not specified, the year fraction 0.5 was accepted.
In a few cases of coincident epochs, a small fraction of a year was voluntarily added or subtracted.
From processing of this data we obtained the value $-0.011 \pm 0.011$ kpc/yr for weighted slope and $+0.006 \pm 0.010$ kpc/yr for unweighted one.

Comparing these data with \citet{Foster2010} we can conclude that our results do not confirm the large trend in the $R_0$ estimates observed by
the authors.
Our computations also revealed a linear trend but much smaller and statistically unreliable, which corresponds to the results of \citet{Surdin1999}
and \citet{Nikiforov2004}.

Another simple but effective test of the significance of the long-term trend in the data (not necessarily linear) is the Abbe criterion.
Strictly speaking, it is intended to test the null hypothesis that all the measurements have the same mathematical expectation.
The Abbe statistics can be computed as the ratio between the Allan variance AVAR (see details of this statistics in \citet{Malkin2008j,Malkin2011ce})
and the variance of the measurements VAR:
\begin{equation}
\begin{array}{rcl}
q &=& \fracd{AVAR}{VAR}\,, \\[1em]
AVAR &=& \fracd{\sum_{i=1}^{n-1}(y_i-y_{i+1})^2}{2(n-1)}\,, \\[1em]
VAR &=& \fracd{\sum_{i=1}^n {(x_i - \bar{x})^2}}{n-1} \,,
\end{array}
\end{equation}
where the mean value $\bar{x}$ is computed by (\ref{eq:uwm_m}).

In presence of significant trend, one can expect that VAR will be substantially greater than AVAR.
In other words, if the value of $q$ is less than corresponding percentage point, we have to reject the null hypothesis, i.e. confess that the input
data contain systematic shift.
In our case, we have obtained $q=1.26$, which is well greater than the 1\% critical value 0.69.

Finally, we can conclude that no statistically significant trend is found in the analyzed data during last 20 years.

\section{Evaluation of $\mathbf{R_0}$ }
\label{sect_compute}

The results $R_0$ determination in 1992--2011 listed in Table~\ref{tab:allr0} were processed with 10 statistical statistical techniques
described in Section~\ref{sect_methods}
The final results obtained by all the methods are presented in Table~\ref{tab:all_results}.

\begin{table}
\centering
\small
\caption{Average estimates of $R_0$ obtained by different statistical methods. The second and third columns contain references to formulas used
for computation of the result given in the last column. See Section~\ref{sect_methods} for method designations}
\label{tab:all_results}
\begin{tabular}{lcc}
\hline
\multicolumn{1}{c}{Method} & $R_0$, kpc & $\chi^2/dof$\\
\hline
WM with $\sigma_{w1}$ & $7.909 \pm 0.051$ & 1.227 \\    
WM with $\sigma_{w2}$ & $7.909 \pm 0.056$ &       \\    
WM with $\sigma_{w3}$ & $7.909 \pm 0.076$ &       \\    
WM with $\sigma_{w4}$ & $7.909 \pm 0.058$ &       \\    
UWM                   & $7.960 \pm 0.062$ &       \\
LRW                   & $7.909 \pm 0.051$ & 1.227 \\
NR                    & $7.909 \pm 0.051$ & 1.227 \\
MP                    & $7.911 \pm 0.059$ & 1.000 \\
M                     & $8.060 \pm 0.075$ &       \\
BM                    & $8.028 \pm 0.083$ &       \\
\hline
\end{tabular}
\end{table}

For MP estimate, we also have got an estimate of the inter-laboratory error $s_b$, which is equal to $0.16$ kpc.
It shows a level of consistency between the input $R_0$ determinations.
Also, this value was added in quadrature to the original uncertainties according to (\ref{eq:mp_weights})
to derive the final MP estimate.

It should be noted that the uncertainty of the LRW, NR, and MP estimates originally computed as $\sigma_{w1}$ can also be computed as
$\sigma_{w2}$, $\sigma_{w3}$ or $\sigma_{w4}$.
We made corresponding calculations, but in our case the results are identical to the WM, and are not presented here.
For more discrepant data these modifications may worth consideration.

\begin{figure}
\centering
\resizebox{0.7\hsize}{!}{\includegraphics[clip]{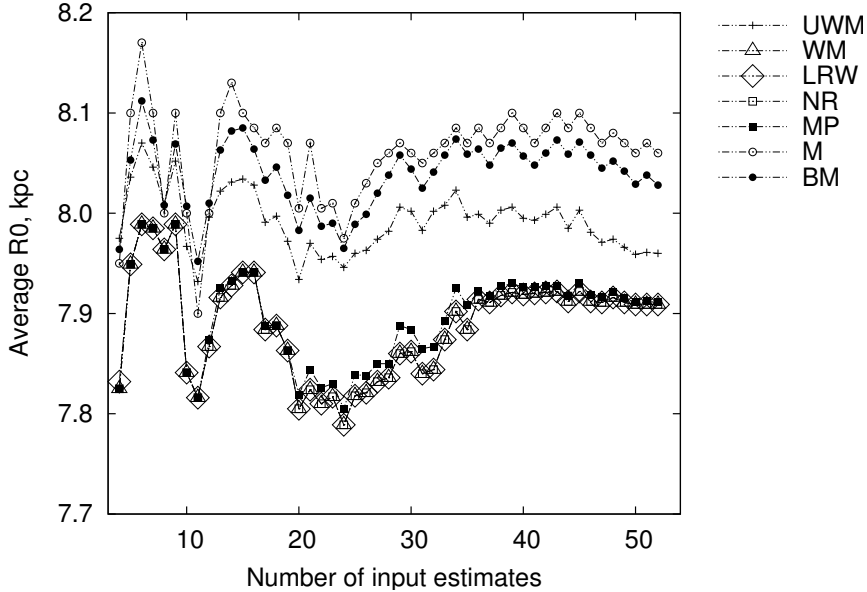}}
\caption{Convergence of $R_0$ estimates. The abscissa axis shows the number of input estimates counted back from the latest one. Unit: kpc.}
\label{fig:convergence}
\end{figure}

Figure~\ref{fig:convergence} shows the convergence of the results obtained with different statistical methods with respect to the number
of input data used in computation of CM.
The data sets used for this computation always begins with the last determination and have different extents in the past.
This test is intended to estimate how much historical data contribute to the current best estimate.
From this test, we can conclude that only the last 35-40 determinations made during the last about 15~years practically define the current
best estimate of $R_0$.

There is no well-founded choice between results obtained with different statistical techniques.
For this reason, it seems reasonable to accept the final $R_0$ value as the middle of the interval overlapping all the estimates of CM
with their 1$\sigma$ uncertainties, namely [7.909-0.076, 8.060+0.075] = [7.833, 8.135].
A half of the span of this interval can be accepted as the uncertainty of the final $R_0$ estimate.
Thus, the following final value of the Galactocentric distance can be accepted:
\begin{equation}
R_0 = 7.98 \pm 0.15 \ \mathrm{kpc}.
\label{eq:final_stat}
\end{equation}

In the framework of the classical method, the value (\ref{eq:final_stat}) considered as a confidence interval corresponds to the confidence probability of 0.995.
However this value of uncertainty is only statistical one indeed.
Various considerations on possible sources and values of systematic errors in determinations of $R_0$ can be found in
\citet{Nishiyama2006,Kerr1986,Reid1993,Nikiforov2004,Groenewegen2008,Nikiforov2008,Trippe2008,Gillessen2009b,Matsunaga2009,Sato2010}.
Analysis of the systematic errors of the published $R_0$ estimates is beyond the scope of this study.
However, it should be mentioned that results of test computations with the original and recomputed $R_0$ estimates suggested by
\citet{Nikiforov2004,Nikiforov2008} did not show any significant impact on the final value.

Taking into account that the $R_0$ measurements analyzed in this study were derived from observations of different objects and making use of several
different methods with different systematic errors, we can presume that these errors are generally independent between methods, and hence are
substantially mitigated in the averaged value (\ref{eq:final_stat}).
Perhaps 0.15--0.2 kpc is a realistic estimate for remaining systematic error in the average result.
On the other hand, a bandwagon effect in the $R_0$ measurements is suspected by many authors, e.g., \citet{Reid1993,Nikiforov2004}
to be substantial.
It may bias the current best estimate, but there is no way to quantify this effect.

Finally, the following value of the Galactocentric distance of the Sun has been obtained in this study:
\begin{equation}
\mathbf{R_0 = 7.98 \pm 0.15\,|_{stat} \pm 0.20\,|_{syst}} \ \mathrm{kpc}.
\label{eq:final}
\end{equation}
which can be considered as the current best estimate of $R_0$.
For most of the practical applications the value of $R_0 = 8.0 \pm 0.25$ kpc can be sufficient.

\section{Concluding remarks}

In this paper we present a new current best estimate of the distance from the Sun to the center of the Galaxy, $R_0$, derived from comparison of
estimates obtained with different statistical techniques.
Some techniques gives mostly ``theoretical'' estimates without accounting for possible discrepancy in the data used.
Other were developed just to evaluate discrepant data and thus should give more realistic estimates.

In spite of $R_0$ estimates used in this study were obtained by different methods and object samples, our analysis has shown that they are rather
consistent than discrepant.
From the results given in Table~\ref{tab:all_results}, one can see that the same or very close values were obtained with WM, LRW, NP, and MP methods.
This is an evidence that there are no large systematic errors in the estimates available in literature.
However, although the estimates derived by several variants of the weighted mean approach give close values of $R_0$, their uncertainties differ
substantially ranging from 0.51 to 0.76 kpc.
Besides, the estimates derived by two variants of median-family technique differ substantially from the WM-family results.
All this means that $R_0$ determinations made during the last 20~years are not fully consistent yet, and using and comparison of different
statistical techniques is justified to obtain a reliable combined $R_0$ estimate.
The same can be said about other astronomical constants derived from combined processing of different determinations.

It is important that our $R_0$ estimate is quite consistent with the values found by \citet{Reid1993} ($8.0 \pm 0.5$ kpc),
\citet{Nikiforov2004} ($7.9 \pm 0.2$ kpc), and \citet{Avedisova2005} ($7.8 \pm\ 0.32$ kpc).
We can conclude from this that the $R_0$ value is to the moment well established around 8.0 kpc with an error of a few tenths of kpc.
This fact, together with absence of a significant trend in the $R_0$ determination made during last 20 years, can lead to a conclusion
that the bandwagon effect may be not so significant as suspected.

Finally, the fast convergence of the $R_0$ determinations found in this study open a possibility of improving $R_0$ uncertainty substantially
during coming years.
It is interesting to notice that the results listed in Table~\ref{tab:means} also show a strong tendency to decreasing the $R_0$ error with time.
Unfortunately, it is mainly explained not by improvement of the observational results, but rather by inhomogeneity of the methods used to estimate
the uncertainty of the average value, as discussed above.

\section*{Acknowledgements}
This research has made extensive use of the SAO/NASA Astrophysics Data System (ADS).

\bibliography{my_eng,galaxy,math}

\begin{thebibliography}{78}
\expandafter\ifx\csname natexlab\endcsname\relax\def\natexlab#1{#1}\fi

\bibitem[{{Alves}(2000)}]{Alves2000}
{Alves}, D.~R. 2000, \apj, 539, 732

\bibitem[{{Ando} {et~al.}(2011){Ando}, {Nagayama}, {Omodaka}, {Handa}, {Imai},
  {Nakagawa}, {Nakanishi}, {Honma}, {Kobayashi}, \& {Miyaji}}]{Ando2011}
{Ando}, K., {Nagayama}, T., {Omodaka}, T., {et~al.} 2011, \pasj, 63, 45

\bibitem[{{Avedisova}(2005)}]{Avedisova2005}
{Avedisova}, V.~S. 2005, \astr, 49, 435

\bibitem[{{Babusiaux} \& {Gilmore}(2005)}]{Babusiaux2005}
{Babusiaux}, C. \& {Gilmore}, G. 2005, \mnras, 358, 1309

\bibitem[{{Bica} {et~al.}(2006){Bica}, {Bonatto}, {Barbuy}, \&
  {Ortolani}}]{Bica2006}
{Bica}, E., {Bonatto}, C., {Barbuy}, B., \& {Ortolani}, S. 2006, \aap, 450, 105

\bibitem[{{Bobylev} {et~al.}(2007){Bobylev}, {Bajkova}, \&
  {Lebedeva}}]{Bobylev2007}
{Bobylev}, V.~V., {Bajkova}, A.~T., \& {Lebedeva}, S.~V. 2007, \astl, 33, 720

\bibitem[{{Bonatto} {et~al.}(2009){Bonatto}, {Bica}, {Barbuy}, \&
  {Ortolani}}]{Bonatto2009}
{Bonatto}, C., {Bica}, E., {Barbuy}, B., \& {Ortolani}, S. 2009, in Globular
  Clusters - Guides to Galaxies, ed. T.~{Richtler} \& S.~{Larsen}, 209--211

\bibitem[{{Carney} {et~al.}(1995){Carney}, {Fulbright}, {Terndrup}, {Suntzeff},
  \& {Walker}}]{Carney1995}
{Carney}, B.~W., {Fulbright}, J.~P., {Terndrup}, D.~M., {Suntzeff}, N.~B., \&
  {Walker}, A.~R. 1995, \aj, 110, 1674

\bibitem[{{Chen} {et~al.}(2011){Chen}, {Geraedts}, {Ouellet}, \&
  {Singh}}]{Chen2011}
{Chen}, J., {Geraedts}, S.~D., {Ouellet}, C., \& {Singh}, B. 2011, Appl. Rad.
  Isot., 69, 1064

\bibitem[{{Cox} \& {Harris}(2004)}]{Cox2004}
{Cox}, M.~G. \& {Harris}, P.~M. 2004, Meas. Tech., 47, 102

\bibitem[{{Dambis} {et~al.}(1995){Dambis}, {Mel'Nik}, \&
  {Rastorguev}}]{Dambis1995}
{Dambis}, A.~K., {Mel'Nik}, A.~M., \& {Rastorguev}, A.~S. 1995, \astl, 21, 291

\bibitem[{{Dambis}(2009)}]{Dambis2009}
{Dambis}, A.~K. 2009, \mnras, 396, 553

\bibitem[{{Dambis}(2010)}]{Dambis2010}
{Dambis}, A.~K. 2010, in Variable Stars, the Galactic halo and Galaxy
  Formation, ed. C.~{Sterken}, N.~{Samus}, \& L.~{Szabados}, 177--180

\bibitem[{{Eisenhauer} {et~al.}(2005){Eisenhauer}, {Genzel}, {Alexander},
  {Abuter}, {Paumard}, {Ott}, {Gilbert}, {Gillessen}, {Horrobin}, {Trippe},
  {Bonnet}, {Dumas}, {Hubin}, {Kaufer}, {Kissler-Patig}, {Monnet},
  {Str{\"o}bele}, {Szeifert}, {Eckart}, {Sch{\"o}del}, \&
  {Zucker}}]{Eisenhauer2005}
{Eisenhauer}, F., {Genzel}, R., {Alexander}, T., {et~al.} 2005, \apj, 628, 246

\bibitem[{{Eisenhauer} {et~al.}(2003){Eisenhauer}, {Sch{\"o}del}, {Genzel},
  {Ott}, {Tecza}, {Abuter}, {Eckart}, \& {Alexander}}]{Eisenhauer2003}
{Eisenhauer}, F., {Sch{\"o}del}, R., {Genzel}, R., {et~al.} 2003, \apjl, 597,
  L121

\bibitem[{{Feast}(1997)}]{Feast1997a}
{Feast}, M.~W. 1997, \mnras, 284, 761

\bibitem[{{Feast} {et~al.}(1998){Feast}, {Pont}, \& {Whitelock}}]{Feast1998}
{Feast}, M., {Pont}, F., \& {Whitelock}, P. 1998, \mnras, 298, L43

\bibitem[{{Feast} \& {Whitelock}(1997)}]{Feast1997c}
{Feast}, M. \& {Whitelock}, P. 1997, \mnras, 291, 683

\bibitem[{{Foster} \& {Cooper}(2010)}]{Foster2010}
{Foster}, T. \& {Cooper}, B. 2010, in \aspc, Vol. 438, The Dynamic Interstellar
  Medium: A Celebration of the Canadian Galactic Plane Survey, ed. R.~{Kothes},
  T.~L. {Landecker}, \& A.~G. {Willis}, 16--30

\bibitem[{{Genzel} {et~al.}(2000){Genzel}, {Pichon}, {Eckart}, {Gerhard}, \&
  {Ott}}]{Genzel2000}
{Genzel}, R., {Pichon}, C., {Eckart}, A., {Gerhard}, O.~E., \& {Ott}, T. 2000,
  \mnras, 317, 348

\bibitem[{{Gerasimenko}(2004)}]{Gerasimenko2004}
{Gerasimenko}, T.~P. 2004, \astr, 48, 103

\bibitem[{{Ghez} {et~al.}(2008){Ghez}, {Salim}, {Weinberg}, {Lu}, {Do}, {Dunn},
  {Matthews}, {Morris}, {Yelda}, {Becklin}, {Kremenek}, {Milosavljevic}, \&
  {Naiman}}]{Ghez2008}
{Ghez}, A.~M., {Salim}, S., {Weinberg}, N.~N., {et~al.} 2008, \apj, 689, 1044

\bibitem[{{Gillessen} {et~al.}(2009{\natexlab{a}}){Gillessen}, {Eisenhauer},
  {Fritz}, {Bartko}, {Dodds-Eden}, {Pfuhl}, {Ott}, \&
  {Genzel}}]{Gillessen2009b}
{Gillessen}, S., {Eisenhauer}, F., {Fritz}, T.~K., {et~al.} 2009{\natexlab{a}},
  \apjl, 707, L114

\bibitem[{{Gillessen} {et~al.}(2009{\natexlab{b}}){Gillessen}, {Eisenhauer},
  {Trippe}, {Alexander}, {Genzel}, {Martins}, \& {Ott}}]{Gillessen2009a}
{Gillessen}, S., {Eisenhauer}, F., {Trippe}, S., {et~al.} 2009{\natexlab{b}},
  \apj, 692, 1075

\bibitem[{{Glass} {et~al.}(1995){Glass}, {Whitelock}, {Catchpole}, \&
  {Feast}}]{Glass1995}
{Glass}, I.~S., {Whitelock}, P.~A., {Catchpole}, R.~M., \& {Feast}, M.~W. 1995,
  \mnras, 273, 383

\bibitem[{{Glushkova} {et~al.}(1998){Glushkova}, {Dambis}, {Mel'Nik}, \&
  {Rastorguev}}]{Glushkova1998}
{Glushkova}, E.~V., {Dambis}, A.~K., {Mel'Nik}, A.~M., \& {Rastorguev}, A.~S.
  1998, \aap, 329, 514

\bibitem[{{Glushkova} {et~al.}(1999){Glushkova}, {Dambis}, \&
  {Rastorguev}}]{Glushkova1999}
{Glushkova}, E.~V., {Dambis}, A.~K., \& {Rastorguev}, A.~S. 1999, \aatr, 18,
  349

\bibitem[{{Gray} {et~al.}(1990){Gray}, {Mac Mahon}, \& {Rajput}}]{Gray1990}
{Gray}, P.~W., {Mac Mahon}, T.~D., \& {Rajput}, M.~U. 1990, Nuclear Instruments
  and Methods in Physics Research A, 286, 569

\bibitem[{{Groenewegen} \& {Blommaert}(2005)}]{Groenewegen2005}
{Groenewegen}, M.~A.~T. \& {Blommaert}, J.~A.~D.~L. 2005, \aap, 443, 143

\bibitem[{{Groenewegen} {et~al.}(2008){Groenewegen}, {Udalski}, \&
  {Bono}}]{Groenewegen2008}
{Groenewegen}, M.~A.~T., {Udalski}, A., \& {Bono}, G. 2008, \aap, 481, 441

\bibitem[{{Gwinn} {et~al.}(1992){Gwinn}, {Moran}, \& {Reid}}]{Gwinn1992}
{Gwinn}, C.~R., {Moran}, J.~M., \& {Reid}, M.~J. 1992, \apj, 393, 149

\bibitem[{{Honma} \& {Sofue}(1996)}]{Honma1996}
{Honma}, M. \& {Sofue}, Y. 1996, \pasj, 48, L103

\bibitem[{{Huterer} {et~al.}(1995){Huterer}, {Sasselov}, \&
  {Schechter}}]{Huterer1995}
{Huterer}, D., {Sasselov}, D.~D., \& {Schechter}, P.~L. 1995, \aj, 110, 2705

\bibitem[{{James} {et~al.}(1992){James}, {Mills}, \& {Weaver}}]{James1992}
{James}, M.~F., {Mills}, R.~W., \& {Weaver}, D.~R. 1992, Nuclear Instruments
  and Methods in Physics Research A, 313, 277

\bibitem[{{Kerr} \& {Lynden-Bell}(1986)}]{Kerr1986}
{Kerr}, F.~J. \& {Lynden-Bell}, D. 1986, \mnras, 221, 1023

\bibitem[{{Layden} {et~al.}(1996){Layden}, {Hanson}, {Hawley}, {Klemola}, \&
  {Hanley}}]{Layden1996}
{Layden}, A.~C., {Hanson}, R.~B., {Hawley}, S.~L., {Klemola}, A.~R., \&
  {Hanley}, C.~J. 1996, \aj, 112, 2110

\bibitem[{{Maciel}(1993)}]{Maciel1993}
{Maciel}, W.~J. 1993, \apss, 206, 285

\bibitem[{{MacMahon} {et~al.}(2004){MacMahon}, {Pearce}, \&
  {Harris}}]{MacMahon2004}
{MacMahon}, D., {Pearce}, A., \& {Harris}, P. 2004, Appl. Rad. Isot., 60, 275

\bibitem[{{Majaess} {et~al.}(2009){Majaess}, {Turner}, \& {Lane}}]{Majaess2009}
{Majaess}, D.~J., {Turner}, D.~G., \& {Lane}, D.~J. 2009, \mnras, 398, 263

\bibitem[{{Majaess}(2010)}]{Majaess2010}
{Majaess}, D. 2010, \actaa, 60, 55

\bibitem[{{Malkin}(2011{\natexlab{a}})}]{Malkin2011fe}
{Malkin}, Z.~M. 2011{\natexlab{a}}, Astron. Rep., 55, 810

\bibitem[{{Malkin}(2011{\natexlab{b}})}]{Malkin2011ce}
{Malkin}, Z.~M. 2011{\natexlab{b}}, \kpcb, 27, 42

\bibitem[{{Malkin}(2001{\natexlab{a}})}]{Malkin2001m}
{Malkin}, Z. 2001{\natexlab{a}}, in 15th Workshop Meeting on European VLBI for
  Geodesy and Astrometry, ed. D.~{Behrend} \& A.~{Rius}, 55--62

\bibitem[{{Malkin}(2001{\natexlab{b}})}]{Malkin2001k}
{Malkin}, Z. 2001{\natexlab{b}}, Communications of the Institute of Applied
  Astronomy RAS, No.~137

\bibitem[{{Malkin}(2008)}]{Malkin2008j}
{Malkin}, Z. 2008, \joge, 82, 325

\bibitem[{{Malkin}(2011{\natexlab{c}})}]{Malkin2011g}
{Malkin}, Z. 2011{\natexlab{c}}, arXiv:1110.6639

\bibitem[{{Matsunaga} {et~al.}(2009){Matsunaga}, {Kawadu}, {Nishiyama},
  {Nagayama}, {Hatano}, {Tamura}, {Glass}, \& {Nagata}}]{Matsunaga2009}
{Matsunaga}, N., {Kawadu}, T., {Nishiyama}, S., {et~al.} 2009, \mnras, 399,
  1709

\bibitem[{{McNamara} {et~al.}(2000){McNamara}, {Madsen}, {Barnes}, \&
  {Ericksen}}]{McNamara2000a}
{McNamara}, D.~H., {Madsen}, J.~B., {Barnes}, J., \& {Ericksen}, B.~F. 2000,
  \pasp, 112, 202

\bibitem[{{Merrifield}(1992)}]{Merrifield1992}
{Merrifield}, M.~R. 1992, \aj, 103, 1552

\bibitem[{{Metzger} {et~al.}(1998){Metzger}, {Caldwell}, \&
  {Schechter}}]{Metzger1998}
{Metzger}, M.~R., {Caldwell}, J.~A.~R., \& {Schechter}, P.~L. 1998, \aj, 115,
  635

\bibitem[{{Moran} {et~al.}(1993){Moran}, {Reid}, \& {Gwinn}}]{Moran1993}
{Moran}, J.~M., {Reid}, M.~J., \& {Gwinn}, C.~R. 1993, in Lecture Notes in
  Physics, Berlin Springer Verlag, Vol. 412, Astrophysical Masers, ed. A.~W.
  {Clegg} \& G.~E. {Nedoluha}, 244--247

\bibitem[{{M\"uller}(1995)}]{Muller1995}
{M\"uller}, J.~W. 1995, Possible advantages of a robust evaluation of
  comparisons. Report BIPM-95/2 (Bureau International des Poids et Mesures,
  S\`evres, France)

\bibitem[{{M\"uller}(2000{\natexlab{a}})}]{Muller2000a}
{M\"uller}, J.~W. 2000{\natexlab{a}}, J. Res. Natl. Inst. Stand. Technol., 105,
  551

\bibitem[{{M\"uller}(2000{\natexlab{b}})}]{Muller2000b}
{M\"uller}, J.~W. 2000{\natexlab{b}}, Weighted median. Report BIPM-2000/6
  (Bureau International des Poids et Mesures, S\`evres, France)

\bibitem[{{Nichols}(2004)}]{Nichols2004}
{Nichols}, A.~L. 2004, Applied radiation and isotopes including data
  instrumentation and methods for use in agriculture industry and medicine, 60,
  247

\bibitem[{{Nikiforov} \& {Petrovskaya}(1994)}]{Nikiforov1994}
{Nikiforov}, I.~I. \& {Petrovskaya}, I.~V. 1994, \astr, 38, 642

\bibitem[{{Nikiforov}(2000)}]{Nikiforov2000}
{Nikiforov}, I.~I. 2000, in \aspc, Vol. 209, IAU Colloq. 174: Small Galaxy
  Groups, ed. M.~J. {Valtonen} \& C.~{Flynn}, 403--407

\bibitem[{{Nikiforov}(2008)}]{Nikiforov2008}
{Nikiforov}, I.~I. 2008, arXiv:0803.0825

\bibitem[{{Nikiforov}(2004)}]{Nikiforov2004}
{Nikiforov}, I. 2004, in \aspc, Vol. 316, Order and Chaos in Stellar and
  Planetary Systems, 199--208

\bibitem[{{Nishiyama} {et~al.}(2006){Nishiyama}, {Nagata}, {Sato}, {Kato},
  {Nagayama}, {Kusakabe}, {Matsunaga}, {Naoi}, {Sugitani}, \&
  {Tamura}}]{Nishiyama2006}
{Nishiyama}, S., {Nagata}, T., {Sato}, S., {et~al.} 2006, \apj, 647, 1093

\bibitem[{{Olling} \& {Merrifield}(1998)}]{Olling1998}
{Olling}, R.~P. \& {Merrifield}, M.~R. 1998, \mnras, 297, 943

\bibitem[{{Paczy\'nski} \& {Stanek}(1998)}]{Paczynski1998}
{Paczy\'nski}, B. \& {Stanek}, K.~Z. 1998, \apjl, 494, L219+

\bibitem[{{Paule} \& {Mandel}(1982)}]{Paule1982}
{Paule}, R. \& {Mandel}, J. 1982, J. Res. Natl. Bur. Stand., 87, 377

\bibitem[{{Pont} {et~al.}(1994){Pont}, {Mayor}, \& {Burki}}]{Pont1994}
{Pont}, F., {Mayor}, M., \& {Burki}, G. 1994, \aap, 285, 415

\bibitem[{{Rastorguev} {et~al.}(1994){Rastorguev}, {Pavlovskaya}, {Durlevich},
  \& {Filippova}}]{Rastorguev1994}
{Rastorguev}, A.~S., {Pavlovskaya}, E.~D., {Durlevich}, O.~V., \& {Filippova},
  A.~A. 1994, \astl, 20, 591

\bibitem[{{Reid} {et~al.}(2009{\natexlab{a}}){Reid}, {Menten}, {Zheng},
  {Brunthaler}, {Moscadelli}, {Xu}, {Zhang}, {Sato}, {Honma}, {Hirota},
  {Hachisuka}, {Choi}, {Moellenbrock}, \& {Bartkiewicz}}]{Reid2009a}
{Reid}, M.~J., {Menten}, K.~M., {Zheng}, X.~W., {et~al.} 2009{\natexlab{a}},
  \apj, 700, 137

\bibitem[{{Reid} {et~al.}(2009{\natexlab{b}}){Reid}, {Menten}, {Zheng},
  {Brunthaler}, \& {Xu}}]{Reid2009b}
{Reid}, M.~J., {Menten}, K.~M., {Zheng}, X.~W., {Brunthaler}, A., \& {Xu}, Y.
  2009{\natexlab{b}}, \apj, 705, 1548

\bibitem[{{Reid}(1989)}]{Reid1989}
{Reid}, M.~J. 1989, in IAU Symposium, Vol. 136, The Center of the Galaxy, ed.
  {M.~Morris}, 37

\bibitem[{{Reid}(1993)}]{Reid1993}
{Reid}, M.~J. 1993, \araa, 31, 345

\bibitem[{{Rukhin} \& {Vangel}(1998)}]{Rukhin1998}
{Rukhin}, A.~L. \& {Vangel}, M.~G. 1998, JASA, 93, 303

\bibitem[{{Sato} {et~al.}(2010){Sato}, {Reid}, {Brunthaler}, \&
  {Menten}}]{Sato2010}
{Sato}, M., {Reid}, M.~J., {Brunthaler}, A., \& {Menten}, K.~M. 2010, \apj,
  720, 1055

\bibitem[{{Shen} \& {Zhu}(2007)}]{Shen2007}
{Shen}, M. \& {Zhu}, Z. 2007, \cjaa, 7, 120

\bibitem[{{Stanek} \& {Garnavich}(1998)}]{Stanek1998}
{Stanek}, K.~Z. \& {Garnavich}, P.~M. 1998, \apjl, 503, L131

\bibitem[{{Stanek} {et~al.}(2000){Stanek}, {Kaluzny}, {Wysocka}, \&
  {Thompson}}]{Stanek2000}
{Stanek}, K.~Z., {Kaluzny}, J., {Wysocka}, A., \& {Thompson}, I. 2000, \actaa,
  50, 191

\bibitem[{{Surdin}(1999)}]{Surdin1999}
{Surdin}, V.~G. 1999, \aatr, 18, 367

\bibitem[{{Trippe} {et~al.}(2008){Trippe}, {Gillessen}, {Gerhard}, {Bartko},
  {Fritz}, {Maness}, {Eisenhauer}, {Martins}, {Ott}, {Dodds-Eden}, \&
  {Genzel}}]{Trippe2008}
{Trippe}, S., {Gillessen}, S., {Gerhard}, O.~E., {et~al.} 2008, \aap, 492, 419

\bibitem[{{Udalski}(1998)}]{Udalski1998}
{Udalski}, A. 1998, \actaa, 48, 113

\bibitem[{{Vanhollebeke} {et~al.}(2009){Vanhollebeke}, {Groenewegen}, \&
  {Girardi}}]{Vanhollebeke2009}
{Vanhollebeke}, E., {Groenewegen}, M.~A.~T., \& {Girardi}, L. 2009, \aap, 498,
  95

\end{thebibliography}
\bibliographystyle{aa}

\end{document}